# Pseudorapidity distributions of the produced charged particles in nucleus-nucleus collisions at low energies on the BNL Relativistic Heavy Ion Collider[*]

JIANG Zhi-Jin (姜志进)[a, †], ZHANG Hai-Li (张海利)[a], and JIANG Guan-Xiang (姜冠祥)[b]

[a]*College of Science, University of Shanghai for Science and Technology, Shanghai 200093, China*

[b]*Institute of Yanke Science and Instrument, Shanghai 200444, China*

**Abstract** The revised Landau hydrodynamic model is used to discuss the pseudorapidity distributions of the produced charged particles in Au+Au and Cu+Cu collisions at energies of $\sqrt{s_{NN}}$ = 19.6 and 22.4 GeV correspondingly on the BNL Relativistic Heavy Ion Collider. It is found that the revised Landau hydrodynamic model alone can give a good description to the experimental measurements. This is different from that in the same collisions but at the maximum energy of $\sqrt{s_{NN}}$ = 200 GeV. Where, in addition to the revised Landau hydrodynamic model, the effects of leading particles have to be taken into account in order to explain the experimental observations. It can be attributed to the different degrees of transparency of participants in different incident energies.

**Key words** Revised Landau hydrodynamic model, produced charged particle, pseudorapidity distribution

## 1 Introduction

Relativistic hydrodynamics provides us a theoretical framework for describing the motion of a continuous flowing medium. It is now widely used to depict various processes for system large to the whole universe and small to the matter created in high energy heavy ion collisions. The experimental observations of the matter produced in collisions, such as the elliptic flow, the single-particle spectra, and the two-particle correlation functions have indeed shown the existence of a collective effect similar to an almost perfect fluid motion [1-4], and can be reasonably well reproduced by hydrodynamic approach. This gives us a confidence to believe that the relativistic hydrodynamics might be one of the best tools for description of the space-time evolution of the matter generated in collisions. Hence, in recent years, the relativistic hydrodynamics has become one of the most active research areas, and has got more and more experimental approvals [5-25].

[*] Supported by the Transformation Project of Science and Technology of Shanghai Baoshan District with Grant No. CXY-2012-25; and by the Shanghai Leading Academic Discipline Project with Grant No. XTKX 2012.

[†] E-mail: Jzj265@163.com

A direct application of the hydrodynamic model is the analysis of the pseudorapidity distributions of the produced charged particles in heavy ion collisions. A wealth of such distributions has been accumulated in experiments [26-31]. In our previous work [6, 7], by taking into account the contributions from leading particles, we have once successfully used the revised Landau hydrodynamic model in describing the experimental measurements carried out by PHOBOS Collaboration in Au+Au and Cu+Cu collisions at the maximum energy of $\sqrt{s_{NN}} = 200$ GeV on RHIC (Relativistic Heavy Ion Collider) at BNL (Brookhaven National Laboratory). Where, the leading particles, as usual, mean the particles which inherit the quantum numbers of colliding nucleons and carry off most part of incident energy. They are then in the large rapidity regions. In Refs. [6, 7], we once argued that these leading particles have the Gaussian rapidity distribution with the normalization constant being equal to the number of participants. This argument is essential in explaining the pseudorapidity distributions of produced charged particles in Au+Au and Cu+Cu collisions at $\sqrt{s_{NN}} = 200$ GeV. Now, what we are concerned is that whether or not the model can still work at low energy. In this paper, we shall use the model to Au+Au and Cu+Cu collisions at energies of $\sqrt{s_{NN}} = 19.6$ and 22.4 GeV correspondingly at BNL-RHIC [26, 27]. We can see that, unlike the cases at energy of $\sqrt{s_{NN}} = 200$ GeV, the revised Landau hydrodynamic model alone can describe well the experimental data at the stated low energies.

**2 Model descriptions**

The revised Landau hydrodynamic model is based on the following assumptions.

(1) The hot and dense matter created in collisions is taken as a massless perfect fluid, which meets the equation of state

$$\varepsilon = 3P , \tag{1}$$

where $\varepsilon$ is the energy density, and $P$ is the pressure. This assumption is now well favored by experimental observations [1-4]. The investigations of the lattice gauge field theory have also shown that the above relation is approximately relevant for the matter with temperature $T > 240$ MeV [32, 33].

(2) During the process of expansion, the fluid quickly achieves local thermal equilibrium. The expansion is adiabatic, and the number of the produced charged particles is proportional to entropy [34, 35]. This means that the entropy in each fluid element or in the whole fluid body is conserved during the hydrodynamic evolution, and the total number of the observed particles can be determined from the initial entropy of the system.

(3) The expansion of fluid undergoes the following two stages [34, 35]. Stage 1: During the fast longitudinal expansion along colliding direction (taken as z axis), there is a simultaneous slow transverse expansion, and the expansions in these two directions advance independently. Stage 2: As the transverse displacement of a fluid element arrives at the initial transverse dimension of the colliding region, the pressure in this fluid element may be neglected. Its rapidity is frozen and therefore remains unchanged. It will have a conic flight with a certain polar angle. The rapidity of the observed particles is determined by that of the fluid element at freeze-out time.

The motion of the fluid observes the equation

$$\frac{\partial T^{\mu\nu}}{\partial x^{\mu}} = 0, \qquad (2)$$

where $x^{\mu} = (x^0, x^1, x^2, x^3) = (t, z, x, y)$ is the space-time 4-vector, and

$$T^{\mu\nu} = (\varepsilon + P)u^{\mu}u^{\nu} - Pg^{\mu\nu} \qquad (3)$$

is the energy-momentum tensor. $u^{\mu}$ and $g^{\mu\nu} = \mathrm{diag}(1,-1,-1,-1)$ are the 4-velocity and metric tensor, respectively.

According to assumption (1) together with above two equations, we can get the expansion equation along longitudinal z direction as

$$\frac{\partial \varepsilon}{\partial t_+} + 2\frac{\partial (\varepsilon e^{-2y})}{\partial t_-} = 0,$$
$$2\frac{\partial (\varepsilon e^{2y})}{\partial t_+} + \frac{\partial \varepsilon}{\partial t_-} = 0, \qquad (4)$$

where $y$ is the rapidity of the fluid element, and

$$t_+ = t + z,$$
$$t_- = t - z,$$

are the light-cone coordinates. The solution of Eq. (4) is

$$\varepsilon(y_+, y_-) = \varepsilon_0 \exp\left[-\frac{4}{3}\left(y_+ + y_- - \sqrt{y_+ y_-}\right)\right], \quad (5)$$

where

$$y_\pm = \ln\left(\frac{\tau}{\Delta_b} e^{\pm y}\right),$$

$\tau$ is the proper time, and

$$\Delta_b = \frac{\sqrt{d^2 - b^2}}{\gamma}$$

is the thickness of the colliding region along $z$ direction for two equal nuclei with diameter $d$ colliding at impact parameter $b$. $\gamma = \sqrt{s_{NN}}/2m_p$ is the Lorentz contract factor, $\sqrt{s_{NN}}/2$ is the center-of-mass energy per nucleon, and $m_p$ is the proton mass.

For slow transverse expansion, it follows the equation

$$\frac{4}{3}\varepsilon u^0 u^0 \frac{\partial v_\phi}{\partial t} = -\frac{\partial P}{\partial \rho}, \quad (6)$$

where $v_\phi$ is the transverse 3-velocity in the direction with the azimuthal angle $\phi$, and $\rho$ is the transverse displacement in this direction. The solution of above equation is

$$\rho(t) = \frac{t^2}{4d_\phi \cosh^2 y}, \quad (7)$$

where $d_\phi$ is the initial distance between the two corresponding points on the boundary of the colliding region at azimuthal angle $\phi$. It is a function of $\phi$ and $b$.

Furthermore, in accordance with assumptions (2) and (3), we can get the rapidity distribution of the produced charged particles as

$$\frac{d^2 N(y, b, \sqrt{s_{NN}}, \phi)}{dy d\phi} = 2cd_\phi \exp\left\{-2\ln(2d_\phi/\Delta_b)\zeta + \sqrt{\left[\ln(2d_\phi/\Delta_b)\zeta\right]^2 - y^2}\right\}, \quad (8)$$

where $c$ is a normalization constant. $\zeta$ is a correction parameter representing the corrections for three factors: the initial configuration of the colliding region, the freeze-out condition, and the assumption of a perfect fluid. For example, in calculations, the initial colliding region is taken as a cylinder with thickness $\Delta_b$, but the reality is that the initial colliding region possesses the shape of an almond being Lorentz contracted along its edge. Furthermore, the freeze-out of the fluid element is supposed to take place as $\rho(t) = d_\phi$. However, the reality may be somewhat different from that [36]. Finally, Eq. (8) is tenable only for a perfect fluid. In realistic case, this equation may have some changes. To take these uncertainties into account, we adopt parameter $\zeta$ to stand for the contributions from them. Since our theoretical knowledge has not advanced to such an extent to determine $\zeta$ in theory, it can now only be fixed by comparing with experimental data.

Eq. (8) shows that the rapidity distribution of the produced charged particles takes on a Gaussian-like form, which is limited in the region of $-\ln(2d_\phi/\Delta_b)\zeta \leq y \leq \ln(2d_\phi/\Delta_b)\zeta$. The value of $\zeta$ influences the region of distribution. The larger is the value of $\zeta$, the broader the rapidity distribution.

Under a certain centrality cut, the value of $\zeta$ is affected by incident energy and nucleus size. For certain incident energy, $\zeta$ decreases with nucleus size. This is due to the fact that the region of rapidity distribution is mainly dependent on incident energy, and is almost independent of nucleus size. Nevertheless, $d_\phi/\Delta_b$ increases with nucleus size. It is evident that, for a given nucleus, $\zeta$ increases with incident energy.

For a given incident energy and nucleus, $\zeta$ increases with centrality cuts. This can be understood if we notice the fact that $d_\phi/\Delta_b$ decreases with the increase of centrality cuts. For example, for Au+Au collisions at $\sqrt{s_{NN}} = 19.6$ GeV, $d_\phi/\Delta_b$ decreases from $\gamma = 10.45$ to 0 for centrality cuts increasing from 0% to 100%. However, the regions of rapidity distribution almost remain unchanged for different centrality cuts.

The total number of the produced charged particles in different azimuthal angles $\phi$ is

$$\frac{dN(y,b,\sqrt{s_{NN}})}{dy} = \int \frac{d^2 N(y,b,\sqrt{s_{NN}},\phi)}{dy d\phi} d\phi. \tag{9}$$

It is a function of rapidity, impact parameter, and beam energy.

## 3. Comparison with experiment data

Having the rapidity distribution of Eq. (9), the pseudorapidity distribution measured in experiments can be expressed as [37]

$$\frac{dN(\eta,b,\sqrt{s_{NN}})}{d\eta} = \sqrt{1 - \frac{m^2}{m_T^2 \cosh^2 y}} \frac{dN(y,b,\sqrt{s_{NN}})}{dy}, \qquad (10)$$

$$y = \frac{1}{2} \ln \left[ \frac{\sqrt{p_T^2 \cosh^2 \eta + m^2} + p_T \sinh \eta}{\sqrt{p_T^2 \cosh^2 \eta + m^2} - p_T \sinh \eta} \right], \qquad (11)$$

where $p_T$ is the transverse momentum, and $m_T = \sqrt{m^2 + p_T^2}$ is the transverse mass.

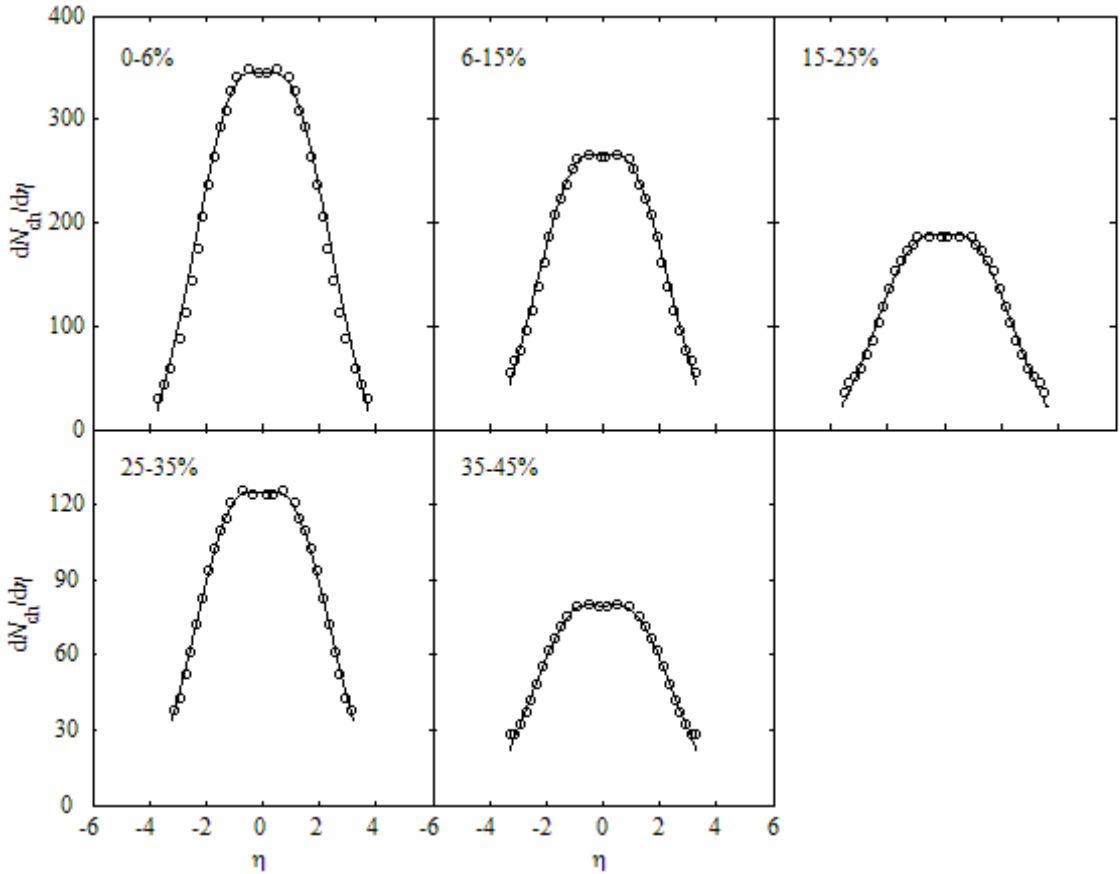

Fig. 1. The pseudorapidity distributions of the produced charged particles in different centrality Au+Au collisions at $\sqrt{s_{NN}} = 19.6$ GeV. The scattered symbols are the experimental measurements [26]. The solid curves are the results obtained from the revised Landau hydrodynamic model of Eq. (9).

Experiments have shown that the overwhelming majority of the produced charged particles in heavy ion collisions at high energy consists of pions, kaons, and protons with proportions of about 83%, 12%, and 5%, respectively [38]. Furthermore, the transverse momentum $p_T$ changes very slowly with centrality cuts. For a specific type of charged particle, it can be well taken as a constant for centrality cuts from 0–55%. This constant is about 0.45, 0.65, and 0.93 GeV/$c$ for pions, kaons, and protons, respectively. In calculations, the $m$ and $p_T$ in Eqs. (10) and (11) take the values of 0.24 GeV and 0.47 GeV/$c$, which are approximately the mean values of those of pions, kaons, and protons.

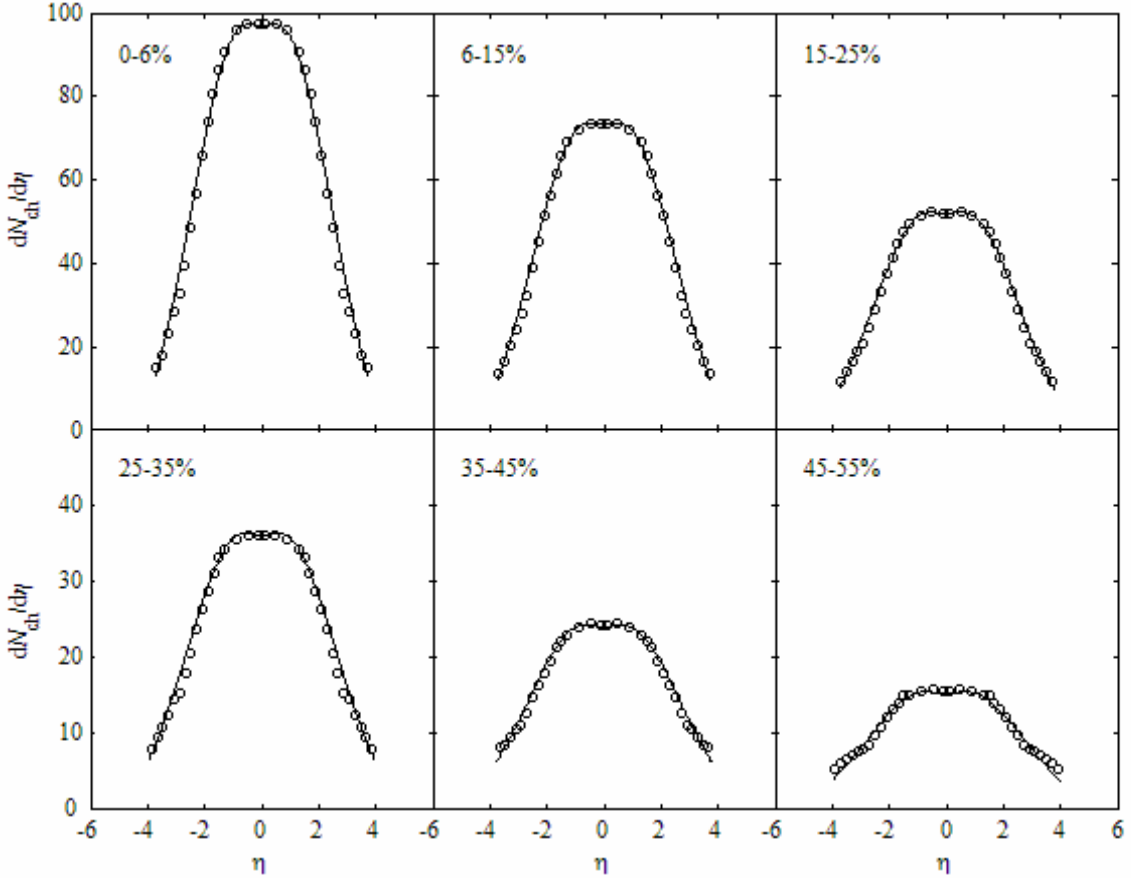

Fig. 2. The pseudorapidity distributions of the produced charged particles in different centrality Cu+Cu collisions at $\sqrt{s_{NN}}=22.4$ GeV. The scattered symbols are the experimental measurements [27]. The solid curves are the results obtained from the revised Landau hydrodynamic model of Eq. (9).

Substituting Eq. (9) into Eq. (10), we can get the pseudorapidity distributions of the produced charged particles. Figures 1 and 2 show such distributions for different centrality Au+Au and Cu+Cu collisions at $\sqrt{s_{NN}}=19.6$ and 22.4 GeV, respectively. The scattered symbols are the experimental measurements [26, 27]. The solid curves are the results from the revised Landau hydrodynamic model of Eq. (9). The corresponding $\chi^2/\text{NDF}$ is 0.1447, 0.0195, and 0.0052 for centrality cuts of 0-6%, 15-25%, and 35-45%, respectively, for Au+Au collisions at $\sqrt{s_{NN}}=19.6$

GeV. For the same centrality cuts in Cu+Cu collisions at $\sqrt{s_{NN}}$ = 22.4 GeV, the $\chi^2/\text{NDF}$ is 0.1013, 0.0421, and 0.0448, respectively. It can be seen that the theoretical results are very consistent with experimental measurements.

In calculations, the correction parameter $\zeta$ in Eq. (8) takes the values of 1.22, 1.23, 1.34, 1.44, and 1.58 for centrality cuts from small to large in Au+Au collisions at $\sqrt{s_{NN}}$ = 19.6 GeV. In Cu+Cu collisions at $\sqrt{s_{NN}}$ = 22.4 GeV, $\zeta$ takes the values of 1.30, 1.41, 1.52, 1.63, 1.76, and 1.92 for centrality cuts from small to large. It can be seen that, for a given nucleus, $\zeta$ increases with centrality cuts, and for a given centrality cut, $\zeta$ decreases with nucleus size. This is consistent with the analyses presented above.

## 4 Summary and discussions

Without considering the effects of leading particles, the revised Landau hydrodynamic model itself can give a good description to the pseudorapidity distributions of the produced charged particles in Au+Au and Cu+Cu collisions at $\sqrt{s_{NN}}$ = 19.6 and 22.4 GeV accordingly at BNL-RHIC. This is different from the collisions at $\sqrt{s_{NN}}$ = 200 GeV, the maximum energy at BNL-RHIC, where the leading particles are essential in explaining experimental measurements. Why doe this difference happen? The answer lies in the degree of transparency of participants in different incident energies. It is known from Ref. [39] that, in Au+Au collisions at $\sqrt{s_{NN}}$ = 200 GeV, the rapidity loss of participants is up to $<\delta y>\approx 2.4$, then the leading particles should locate at

$$y_0 = y_{beam} - <\delta y> = 5.36 - 2.40 = 2.96,$$

which is close to the fitting parameter of $y_0 = 2.75$ in our previous work [6]. Hence, in collisions at $\sqrt{s_{NN}}$ = 200 GeV, the mid-rapidity region is nearly net baryon-free, or the participants are almost transparent. On the other hand, for collisions at low energies, the rapidity loss is about $<\delta y>\approx 0.58 y_{beam}$. Thus the leading particles should locate at

$$y_0 = y_\text{beam} - <\delta y> = 0.42 y_\text{beam} = \begin{cases} 1.28, \left(\sqrt{s_{NN}} = 19.6 \text{ GeV}\right), \\ 1.33, \left(\sqrt{s_{NN}} = 22.4 \text{ GeV}\right). \end{cases}$$

It is so close to the mid-rapidity that the leading particle effect is hidden by the large yield of charged particles. Hence, in collisions at low energies, the mid-rapidity region is high-baryon dense, or the participants are almost full stopping.

## References

[1] Ollitrault J Y. Phys. Rev. D, 1992, **46**: 229-245

[2] Adler S S et al (PHENIX Collaboration). Phys. Rev. Lett., 2003, **91**: 182301

[3] Aamodt K et al (ALICE Collaboration). Phys. Rev. Lett., 2011, **107**: 032301

[4] Chatrchyan S et al (CMS Collaboration). Phys. Rev. C, 2013, **87**: 014902

[5] Wong C Y. Phys. Rev. C, 2008, **78**: 054902

[6] Jiang Z J, Li Q G, and Zhang H L. Phys. Rev. C, 2013, **87**: 044902

[7] Jiang Z J, Li Q G, and Zhang H L. Nucl. Phys. Rev. 2013, **30**: 26-31 (in Chinese)

[8] Lou X H, Jiang Z J, and Li Q G. Nucl. Phys. Rev. 2012, **29**: 52-56 (in Chinese)

[9] Jiang Z J and Sun Y F. Nucl. Phys. Rev. 2010, **27**: 421-425 (in Chinese)

[10] Jiang Z J, Sun Y F, and Ni W X. Chinese Phys. C, 2010, **34**: 1104-1110

[11] Wang Z W and Jiang Z J. Chinese Phys. C, 2009, **33**: 274-280

[12] Dong Y F, Jiang Z J, and Wang Z W. Chinese Phys. C, 2008, **32**: 259-263

[13] Csörgő T, Nagy M I, and Csanád M. Phys. Lett. B, 2008, **663**: 306-311

[14] Nagy M I, Csörgő T, and Csanád M. Phys. Rev. C, 2008, **77**: 024908

[15] Csanád M, Nagy M I, and Lökös S. EPJ A, 2012, **48**: 173-178

[16] Sarkisyan E K G and Sakharov A S. Eur. Phys. J. C, 2010, **70**: 533-541

[17] Gale C, Jeon S, and Schenke B. Intern. J. Mod. Phys. A, 2013, **28**: 1340011

[18] Bialas A, Janik R A, and Peschanski R. Phys .Rev. C, 2007, **76**: 054901

[19] Bialas A and Peschanski R. Phys. Rev. C, 2011, **83**: 054905

[20] Borshch M S and Zhdanov V I. SIGMA, 2007, **3**: 116-126

[21] Steinberg P A. Nucl. Phys. A, 2005, **752**: 423-432

[22] Schenke B, Jeon S, and Gale C. Phys. Rev. C, 2012, **85**: 024901

[23] Nonaka C and Bass Steffen A. Phys. Rev. C, 2007, **75**: 014902

[24] Shen C, Heinz U, Huovinen P, and Song H. Phys. Rev. C, 2010, **82**: 054904

[25] Song H, Bass Steffen A, Heinz U, Hirano T, and Shen C. Phys. Rev. Lett., 2011, **106**: 192301

[26] Back B B et al (PHOBOS Collaboration). Nucl. Phys. A, 2005, **757**: 28-101

[27] Alver B et al (PHOBOS Collaboration). Phys. Rev. Lett., 2009, **102**: 142301

[28] Back B B et al (PHOBOS Collaboration). Phys. Rev. Lett., 2005, **94**: 082304

[29] Antchev G et al (TOTEM Collaboration). Eur. phys. Lett., 2012, **98**: 31002

[30] Khachatryan V et al (CMS Collaboration). Phys. Rev. Lett., 2010, **105**: 022002

[31] Toia A et al (ALICE Collaboration). J. Phys. G: Nucl. Part. Phys., 2011, **38**: 124007

[32] Borsányi S, Endrődi G, Fodor Z, Jakovác A, Katz S D, Krieg S, Ratti C, and Szabó K K. JHEP, 2010, **77**: 1-31

[33] Adare A et al (PHENIX Collaboration). Phys. Rev. Lett., 2007, **98**: 162301.

[34] Landau L D. Izv. Akad. Nauk SSSR, 1953, **17**: 51-64 (in Russian)

[35] Belenkij S Z and Landau L D. Uspekhi Fiz. Nauk, 1955, **56**: 309-348 (in Russian)

[36] Cooper F and Frye G. Phys. Rev. D, 1974, **10**: 186-189

[37] Wong C Y. Introduction to high energy heavy ion collisions, Press of Harbin Technology University, Harbin, China, 2002, p 23-45 (in Chinese)

[38] Adler S S et al (PHENIX Collaboration). Phys. Rev. C, 2004, **69**: 034909

[39] Bearden I G et al (BRAHMAS Collaboration). Phys. Rev. Lett., 2004, **93**: 102301